# In situ calibration of camera-refraction interface based on analytical refractive imaging equation

HONGZHE WANG,[1] YANG SONG,[1,*] HUAJUN CAI,[1] LIN BO,[1] BOYAN ZHANG,[1] YUNJING JI,[1] JIANCHENG LAI,[1] AND ZHENHUA LI[1]

[1]*Department of Information Physics and Engineering, Nanjing University of Science and Technology, Nanjing 210094, China*

**sy0204@njust.edu.cn*

**Abstract:** Camera calibration is an essential process in photogrammetry, serving as a crucial link between the 2D image coordinate system and the 3D world coordinate system. However, when observations are conducted through refractive interfaces, the refraction effects at these interfaces render traditional calibration methods ineffective, significantly compromising measurement accuracy. To address this challenge, we propose a novel camera calibration method based on the analytical refractive imaging (ARI) equation. The ARI method facilitates accurate estimation of camera parameters from distorted images and enables in-situ joint calibration of both the camera and the refractive interface. The experimental results indicate that the proposed method reduces the error to only 10% of that produced by conventional ray-tracing (RT) method. Moreover, while maintaining comparable computational accuracy and efficiency, it effectively mitigates the local convergence issues that may arise in the polynomial fitting (PF) approach. Finally, reconstruction experiments further confirm the accuracy of the proposed method. Experimental results demonstrate that the proposed method outperforms existing refractive calibration techniques in terms of accuracy while maintaining high precision in 3D reconstruction tasks.

## 1. Introduction

Due to its advantages of non-contact measurement and high precision, photogrammetry technology has been widely applied in various fields. However, in many application scenarios, such as three-dimensional reconstruction of confined flames in combustion chambers [1-3], flow field detection in supersonic wind tunnels [4-6], and underwater visual measurement in ocean engineering [7-9], a refractive interface exists between the camera and the object to be measured for the purpose of protecting the camera or setting up the scene. In these scenarios, traditional camera calibration methods are no longer applicable, and high-precision camera calibration through refractive interfaces has become a pressing problem to be addressed.

Several studies have been conducted on the refractive correction of refractive interfaces, and the related methods can be mainly divided into two categories: image calibration and geometric calibration. Image calibration mainly refers to treating the influence of refraction on images as a type of distortion, and solving distortion parameters to correct the images, making them close to images unaffected by refraction [10]. For example, Kang et al. [11] employed a single viewpoint (SVP) model that included radial lens distortion correction and focal length adjustment for the three-dimensional reconstruction of underwater objects. However, this method cannot completely solve the refraction problem nor reveal the essence of refractive imaging. Teibiz et al. [12] analyzed the applicability of the SVP model in refraction problems and proved that the SVP model is invalid for such refractive vision systems. Subsequently, they simulated images unaffected by refraction by changing the lens settings. Nevertheless, this process is based on the assumption that the refractive interface is parallel to the camera imaging plane, which is difficult to achieve in practical operations

The SVP model fails to accurately reflect refraction, and a simpler and more straightforward approach is to use polynomial fitting methods. Haile et al. [13] constructed a polynomial mapping function using image registration technology to correct underwater refractive images; however, this method requires an unrefracted reference image as a benchmark. Samper et al. [14] directly corrected refractive images using a six-degree-of-freedom polynomial function and successfully applied it to 3D reconstruction of the sole under a glass interface, though this approach often requires multiple correction planes to build complex polynomial combinations. Polynomial models have also been widely used in PIV/PTV technologies [15-17] to calculate velocity field displacements under refraction conditions. Although this method establishes a mapping relationship between 3D object points and 2D image points through a calibration procedure and can adapt to complex refractive geometries, it inherently has critical defects: the model parameters lack clear physical meanings [18], making them difficult to interpret physically. It can only achieve high precision within the calibration area, while severe performance degradation occurs outside the area.

Compared with image correction methods, geometric correction methods place greater emphasis on the physical interpretability of the model. Their core principle is to combine the perspective imaging model with Snell's law for ray tracing to accurately calculate the actual imaging position of object points. Sebastian et al. [19] proposed a Refractive Correction Ray Tracing (RCRT) algorithm, which was successfully applied to reconstructing seahorses underwater. Wang et al. [20] developed a non-single-viewpoint ray-tracing model for calibrating underwater structured-light 3D reconstruction systems, though the camera model still relies on the traditional SVP (Single Viewpoint) assumption. Mulsow [21,22] proposed improved methods for multi-refractive interfaces, solving interface intersections via backward ray tracing (BRT) combined with physical constraints. To overcome the problem of non-physical paths potentially generated by BRT, an alternating forward ray tracing (AFRT) algorithm was further developed, which uses bundle adjustment for iterative solving until convergence. But there are some drawbacks, such as slow convergence or even non-convergence. Belden [23] improved the AFRT algorithm, developing efficient calibration models for planar and cylindrical interfaces in fluid mechanics experiments. Robin et al. [24] proposed an optimization method that significantly improves computational efficiency while maintaining accuracy, suitable for both planar and cylindrical refraction scenarios.

Geometric correction methods can also be combined with triangulation and multi-camera vision to optimize for dual-camera and multi-camera problems. In the case of photographing underwater objects, Sun et al. [25, 26] employed a ray-tracing method to establish correspondence between images from two cameras, thereby improving the efficiency of feature point matching in underwater images. They also applied a differential evolution algorithm to optimize global parameters, achieving underwater reconstruction of propeller structures. To perform multi-camera calibration, Feng et al. [27] used a transparent glass calibration plate, where half of the cameras needed to acquire calibration points through the glass. They also adopted ray tracing as a solution and optimized global parameters via bundle adjustment.

In previous work, we derived the ARI equation [28], whose form is shown in Eq. (1). This equation clearly describes the process of a 3D point being projected onto a 2D pixel plane through a refractive interface. Building on this, this paper proposes a camera calibration method that incorporates the refractive interface.

$$s'\tilde{\mathbf{p}} = \mathbf{K}\mathbf{R}_{\mathrm{IC}}\mathbf{X}_{\mathrm{R}}\begin{bmatrix}\mathbf{R}_{\mathrm{WI}} & \mathbf{t}_{\mathrm{WI}} \\ \mathbf{0}^{\mathrm{T}} & 1\end{bmatrix}\tilde{\mathbf{P}}_{\mathrm{W}} \tag{1}$$

Where s' is a factor used for normalization, $\tilde{\mathbf{p}} = [u, v, 1]^T$ is the 2D homogeneous coordinate in the pixel coordinate system, K is the intrinsic matrix of camera, $\mathbf{X_R}$ is a transformation matrix derived from the ray transmission matrix, $\tilde{\mathbf{P}}_{\mathrm{W}} = [x_{\mathrm{W}}^P, y_{\mathrm{W}}^P, z_{\mathrm{W}}^P, 1]^{\mathrm{T}}$ is the 3D homogeneous coordinate of a point P in the world coordinate system, $\mathbf{R}_{\mathrm{IC}}$ is the rotation matrix; and $\mathbf{t}_{\mathrm{IC}}$ is the translation vector, which transforms from refractive interface coordinate system to camera coordinate system, $\mathbf{R}_{\mathrm{WI}}$ and $\mathbf{t}_{\mathrm{WI}}$ are the rotation matrix and the translation vector which transform from the world coordinate system to refractive interface coordinate system. The subsequent symbols in this article are also named according to this rule.

The structure of this article is organized as follows: Chapter 2 elaborates on the theoretical derivation process of the extension center, proposes an analytical solution method for each parameter in the ARI equation, and introduces a beam adjustment algorithm aimed at achieving global optimization of these parameters. Chapter 3 constructs a comprehensive experimental verification platform and systematically evaluates the performance differences between the proposed method and traditional ray tracing and polynomial fitting methods, focusing on calibration accuracy and stability through comparative experiments. Chapter 4 implements calibration for a multi-camera system based on a rotating calibration plate scheme, while analyzing the accuracy degradation issue associated with the polynomial fitting method outside its fitting area. Chapter 5 validates the actual accuracy and reliability of the proposed calibration method at an application level through three-dimensional reconstruction experiments. Finally, Chapter 6 provides a comprehensive summary of the entire research work conducted.

## 2. Purpose of refractive imaging calibration method

Corresponding to Eq (1), the expansion form of the ARI equation is shown in Eq. (2).

$$s'\tilde{\mathbf{p}} = \begin{bmatrix} f_u & 0 & u_0 \\ 0 & f_v & v_0 \\ 0 & 0 & 1 \end{bmatrix} \begin{bmatrix} \omega(r_y - r_x r_z) & \omega(-r_y - r_x r_z) & -r_x \\ \omega(-r_x - r_y r_z) & \omega(r_x - r_y r_z) & -r_y \\ \omega(1 - r_z^2) & \omega(1 - r_z^2) & -r_z \end{bmatrix} \begin{bmatrix} 1 & 0 & 0 & 0 \\ 0 & 1 & 0 & 0 \\ 0 & 0 & 1 & \left(d - d/n\right) \end{bmatrix} \begin{bmatrix} \mathbf{R}_{\mathrm{WI}} & \mathbf{t}_{\mathrm{WI}} \\ \mathbf{0}^{\mathrm{T}} & 1 \end{bmatrix} \tilde{\mathbf{P}}_{\mathrm{W}} \quad (2)$$

Where $f_u$, $f_v$, $u_0$ and $v_0$ are internal parameters of the camera; $r_x$, $r_y$, and $r_z$ are the normal direction components of the refractive interface, $\omega = 1/\sqrt{2(1 - r_z^2)}$; $d$ and $n$ represent the thickness and refractive index of the refractive interface.

The objective of calibration is to determine the parameters in the ARI equations, thereby establishing a mapping relationship between the 3D world coordinate system and the 2D image coordinate system. To simplify the formulation, the matrix in Eq. (2) is transformed as shown in Eq. (3).

$$s'\tilde{\mathbf{p}} = \mathbf{K}\mathbf{R}_{\mathrm{IC}}\mathbf{X}_{\mathrm{R}}\mathbf{T}_{\mathrm{WI}}\tilde{\mathbf{P}}_{\mathrm{W}} \quad (3)$$

$\mathbf{R}_{\mathrm{IC}}$ represents the rotation matrix for converting the interface coordinate system to the camera coordinate system, and $\mathbf{T}_{\mathrm{WI}}$ represents the rigid transformation matrix for converting the world coordinate system to the interface coordinate system. The subsequent sections will systematically derive the solutions for each matrix.

### *2.1 Determination of the matrix $\mathbf{R}_{\mathrm{IC}}$ for achieving interface orientation*

The determination of matrix $\mathbf{R}_{\mathrm{IC}}$ is essentially equivalent to determining the orientation of the refraction interface. Due to the fact that the orientation of the refractive interface has different impacts on camera imaging, the orientation of the refractive interface must be considered together during calibration. The direction of the emergent ray remains unchanged for the light incident perpendicularly to the refractive interface (purple line in Fig. 1), and its imaging position $O'$ is defined as the expansion center with coordinates $(u_s, v_s)$. The coordinates of the expansion center satisfy the relationship shown in Eq. (4) [28], Gong et al. [29] also proposed a similar theory.

$$\begin{cases} u_s = u_0 + f_u \dfrac{r_x}{r_z} \\ v_s = v_0 + f_v \dfrac{r_y}{r_z} \end{cases} \tag{4}$$

Note that vector $\mathbf{n} = [r_x,\ r_y,\ r_z]^T$ is a unit vector, we can obtain the result shown in Eq. (5).

$$\begin{cases} r_x = \dfrac{f_v(u_s - u_0)}{\sqrt{f_u^2(v_s - v_0)^2 + f_v^2(u_s - u_0)^2 + f_u^2 f_v^2}} \\ r_y = \dfrac{f_u(v_s - v_0)}{\sqrt{f_u^2(v_s - v_0)^2 + f_v^2(u_s - u_0)^2 + f_u^2 f_v^2}} \\ r_z = \dfrac{f_u f_v}{\sqrt{f_u^2(v_s - v_0)^2 + f_v^2(u_s - u_0)^2 + f_u^2 f_v^2}} \end{cases} \tag{5}$$

Therefore, if the coordinates of the expansion center are known, the direction of the refractive interface can be calculated. To obtain the expansion center coordinates, we proposed the method of stacking glass, whose principle is shown in Fig. 1.

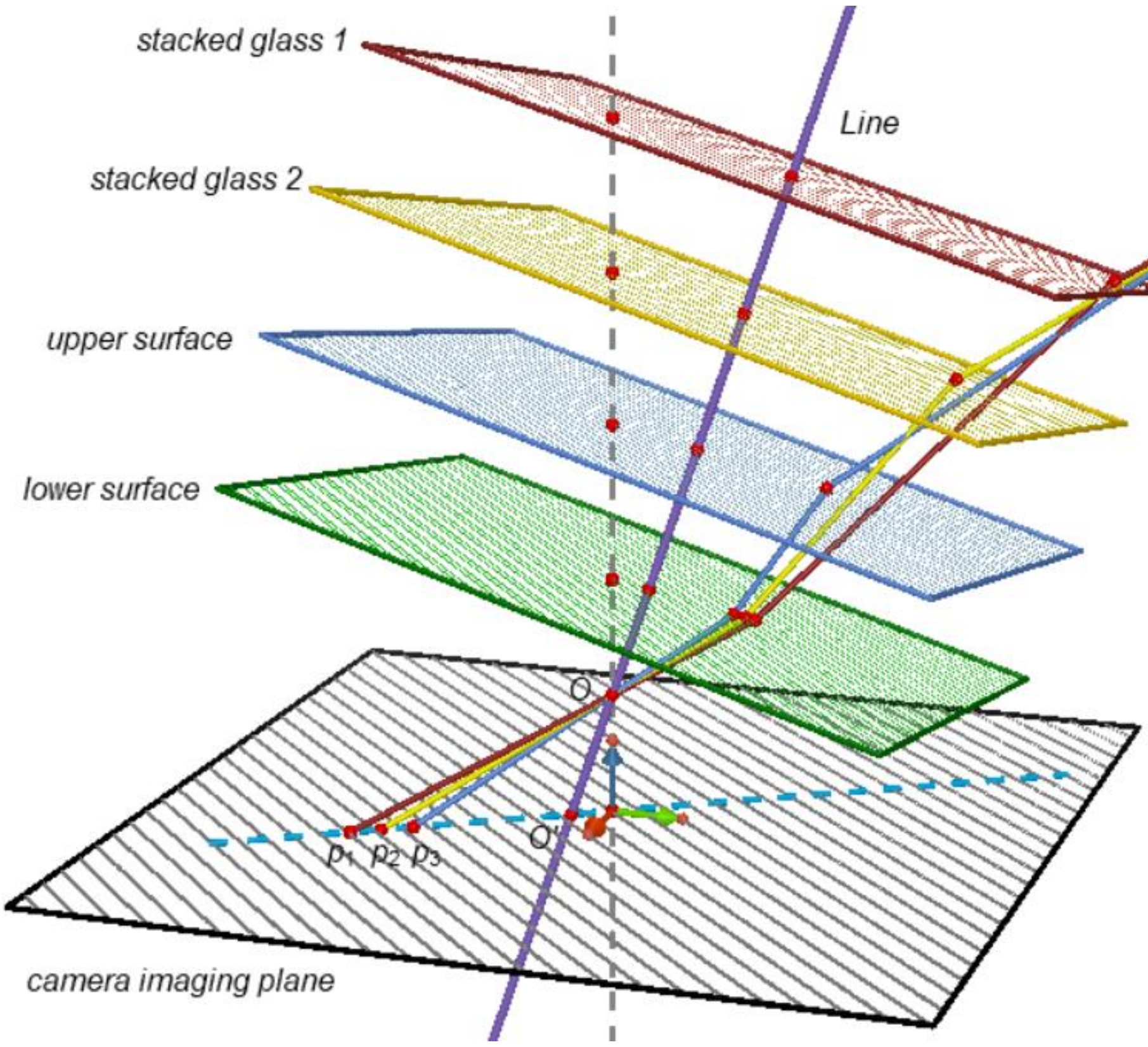

Fig. 1. The influence of glass with different thicknesses on imaging. (The red, yellow and blue planes correspond to the upper surfaces of three glass Windows of different thicknesses, the green plane represents the lower surface shared by these glass Windows, and the black plane represents the imaging plane of the camera.)

As shown in Fig. 1, This study simulated the imaging process of three different thicknesses of glass Windows. The green plane represents the lower surface of the glass window, while the blue plane denotes its upper surface. The red and yellow planes correspond to scenarios where two distinct glass thicknesses are stacked atop the window. When a 3D spatial point P is projected through these glass media onto the image plane, its light path strictly adheres to Snell's law, with all rays passing through the camera's optical center O. The light trajectories for the three thickness conditions are color-coded in red, yellow, and blue, matching their respective upper refractive interfaces, and their corresponding image points are labeled $p_1$, $p_2$ and $p_3$.

The results demonstrate that variations in refractive interface thickness induce positional shifts in the image points. According to the fundamental principle of refraction, incident and refracted rays are coplanar. Consequently, points $p_1$, $p_2$, $p_3$, and the extended center O' must be collinear—a geometric relationship that holds universally for all object points. Leveraging this property, when imaging trajectories of multiple object points are available, the coordinates of the extended center can be precisely determined by fitting the intersection of these trajectories.

The proposed method enables efficient calibration of refractive surface orientation. Once the optical system is fixed, in-situ calibration can be performed directly without requiring disassembly or readjustment. Unlike numerical iteration approaches, this method establishes a physical model based on actual optical measurement processes, fundamentally avoiding convergence issues that may arise with iterative algorithms.

### *2.2 Determination of the matrix* $\mathbf{X}_\mathrm{R}$

The determination of matrix $\mathbf{X}_\mathrm{R}$ is essentially equivalent to determining the thickness and refractive index of the refractive interface. In a typical experimental setup, the thickness and material parameters of the observation window are usually determined during the manufacturing stage and can thus be treated as known quantities. When these parameters are unknown, direct measurement methods such as vernier calipers or interferometers can be employed for precise determination.

The direct measurement approach generally outperforms numerical iteration methods in terms of parameter acquisition accuracy: the former directly reflects the true physical characteristics of the actual object, while the latter is constrained by the local convergence properties of optimization algorithms, potentially leading to systematic deviations between the obtained solutions and the true physical parameters.

### *2.3 Determination of the matrix* $\mathbf{K}$

We define the product of the three middle matrices on the right side of eq. (1) as the generalized extrinsic matrix $[\mathbf{R} \quad \mathbf{t}]$, as shown in Eq. (6).

$$[\mathbf{R} \quad \mathbf{t}] = \mathbf{R}_\mathrm{IC}\mathbf{X}_\mathrm{R}\begin{bmatrix} \mathbf{R}_\mathrm{WI} & \mathbf{t}_\mathrm{WI} \\ \mathbf{0}^\mathrm{T} & 1 \end{bmatrix} \tag{6}$$

At this point, Eq. (1) can be transformed into the form shown in Eq. (7).

$$s'\tilde{\mathbf{p}} = \mathbf{K}[\mathbf{R} \quad \mathbf{t}]\tilde{\mathbf{P}}_\mathrm{W} \tag{7}$$

When the matrix $\mathbf{R}$ satisfies the condition of the unit orthogonal matrix, the system can be transformed into the traditional camera calibration framework. At this time, the camera's internal parameter matrix $\mathbf{K}$ and the generalized external parameter matrix $[\mathbf{R} \quad \mathbf{t}]$ can be solved by the classical calibration method [30]. The unit orthogonality of matrix $\mathbf{R}$ is proved as follows:

According to Eq. (2), the matrix $\mathbf{X}_{\mathrm{R}}$ can be decomposed into a linear combination of the identity matrix and the translation vector. Based on this, Eq. (6) can be expressed in the following form:

$$\begin{bmatrix}\mathbf{R} & \mathbf{t}\end{bmatrix}=\mathbf{R}_{\mathrm{IC}}\begin{bmatrix}\mathbf{I} & \mathbf{t}_{\mathrm{R}}\end{bmatrix}\begin{bmatrix}\mathbf{R}_{\mathrm{WI}} & \mathbf{t}_{\mathrm{WI}} \\ \mathbf{0}^{\mathrm{T}} & 1\end{bmatrix} \tag{8}$$

Where $\mathbf{I}$ is the identity matrix. Multiplying the matrices on the right side of the equation in sequence gives the result:

$$\begin{bmatrix}\mathbf{R} & \mathbf{t}\end{bmatrix}=\begin{bmatrix}\mathbf{R}_{\mathrm{IC}} & \mathbf{R}_{\mathrm{IC}}\mathbf{t}_{\mathrm{R}}\end{bmatrix}\begin{bmatrix}\mathbf{R}_{\mathrm{WI}} & \mathbf{t}_{\mathrm{WI}} \\ \mathbf{0}^{\mathrm{T}} & 1\end{bmatrix} \tag{9}$$

$$\begin{bmatrix}\mathbf{R} & \mathbf{t}\end{bmatrix}=\begin{bmatrix}\mathbf{R}_{\mathrm{IC}}\mathbf{R}_{\mathrm{WI}} & \mathbf{R}_{\mathrm{IC}}\mathbf{t}_{\mathrm{WI}}+\mathbf{R}_{\mathrm{IC}}\mathbf{t}_{\mathrm{R}}\end{bmatrix} \tag{10}$$

Therefore, $\mathbf{R}=\mathbf{R}_{\mathrm{IC}}\mathbf{R}_{\mathrm{WC}}$, $\mathbf{R}_{\mathrm{IC}}$ and $\mathbf{R}_{\mathrm{WC}}$ are rotation matrices between different coordinate systems, satisfying the property of unit orthogonality. Therefore, $\mathbf{R}$ is also a unit orthogonality matrix. The initial solutions of the camera's internal parameter matrix $\mathbf{K}$ and the generalized external parameter matrix $[\mathbf{R} \quad \mathbf{t}]$ can be solved by the classical calibration method

### *2.4 Determination of the matrix* $\mathbf{T}_{\mathrm{WI}}$

At this juncture, we have successfully derived $[\mathbf{R} \quad \mathbf{t}]$, $\mathbf{R}_{\mathrm{IC}}$ and $\mathbf{X}_{\mathrm{R}}$, as presented in Eq. (6). Once these matrices are determined, they can be written in the form of a coefficient matrix, as shown in Eq. (11). The elements in $\mathbf{A}_{3\times 4}$ are determined by the direction, thickness and refractive index of the refractive interface, while $\mathbf{B}_{3\times 4}$ is determined by the generalized external parameter matrix.

$$\mathbf{B}_{3\times 4}=\mathbf{A}_{3\times 4}\begin{bmatrix}\mathbf{R}_{\mathrm{WI}} & \mathbf{t}_{\mathrm{WI}} \\ \mathbf{0}^{\mathrm{T}} & 1\end{bmatrix} \tag{11}$$

$$\begin{bmatrix}
a_{11} & a_{12} & a_{13} & & & & & & & & & & \\
 & & & a_{11} & a_{12} & a_{13} & & & & & & & \\
 & & & & & & a_{11} & a_{12} & a_{13} & & & & \\
 & & & & & & & & & a_{11} & a_{12} & a_{13} & a_{14} \\
a_{21} & a_{22} & a_{23} & & & & & & & & & & \\
 & & & a_{21} & a_{22} & a_{23} & & & & & & & \\
 & & & & & & a_{21} & a_{22} & a_{23} & & & & \\
 & & & & & & & & & a_{21} & a_{22} & a_{23} & a_{24} \\
a_{31} & a_{32} & a_{33} & & & & & & & & & & \\
 & & & a_{31} & a_{32} & a_{33} & & & & & & & \\
 & & & & & & a_{31} & a_{32} & a_{33} & & & & \\
 & & & & & & & & & a_{31} & a_{32} & a_{33} & a_{34}
\end{bmatrix}
\begin{bmatrix} r_{11} \\ r_{21} \\ r_{31} \\ r_{12} \\ r_{22} \\ r_{32} \\ r_{13} \\ r_{23} \\ r_{33} \\ t_x \\ t_y \\ t_z \\ 1 \end{bmatrix}
=
\begin{bmatrix} b_{11} \\ b_{12} \\ b_{13} \\ b_{14} \\ b_{21} \\ b_{22} \\ b_{23} \\ b_{24} \\ b_{31} \\ b_{32} \\ b_{33} \\ b_{34} \end{bmatrix} \tag{12}$$

Eq. (11) can be expanded to obtain the well-posed system of equations as shown in Eq. (12), so each element in $[\mathbf{R}_{\mathrm{WI}} \quad \mathbf{t}_{\mathrm{WI}}]$ can be solved and has a unique solution. Where $a_{ij}$, $b_{ij}$ and $r_{ij}$ are the elements in matrices $\mathbf{A}_{3\times 4}$, $\mathbf{B}_{3\times 4}$ and $\mathbf{R}_{\mathrm{WI}}$, respectively. $t_x$, $t_y$, $t_z$ are the elements in vector $\mathbf{t}_{\mathrm{WI}}$.

### *2.5 Optimize global parameters through bundle adjustment*

Through the above calibration method, all unknown parameters in Eq. (1), which is ARI equation, can be solved, so as to establish the mapping relationship between the 3D space points refracted by glass and the 2D pixel coordinates. However, due to inevitable noise interference and error accumulation during the measurement process, their values are often insufficiently accurate. Therefore, in the final step of calibration, the bundle adjustment algorithm is used to globally optimize these parameters. The bundle adjustment algorithm [31-33] improves accuracy by minimizing the sum of reprojection errors of all image points. The Levenberg-Marquardt (LM) algorithm [34-36] is the most commonly used, which combines the gradient descent method and the Gauss-Newton method. By dynamically adjusting the damping factor, it balances the convergence speed and stability, enabling fast convergence in large-scale initial estimates. The cost function is given by the following equation:

$$\min_{\mathbf{N},\mathbf{K},\mathbf{R},\mathbf{t}} f = \sum_{i=1}^{a} \left\| m_i - \tilde{m}_i(\mathbf{N},\mathbf{K},\mathbf{R},\mathbf{t}) \right\|^2 \tag{13}$$

Where $a$ is the sum of feature points captured by cameras, $\mathbf{N}$ represents the normal vector of the refractive interface, $\mathbf{K}$ is the camera intrinsic matrix, $\mathbf{R}$ is the camera rotation matrix, $\mathbf{t}$ is the camera translation vector, and $m$ and $\tilde{m}$ are the pixel coordinates from actual images and the 2D projected coordinates of 3D points via the parameters mentioned earlier, respectively. Eq. (13) is optimized using bundle adjustment based on the LM algorithm to obtain the optimal parameters for all cameras and the refractive plane.

## 3. Calibration accuracy comparison experiment

### *3.1 Experimental configuration with one camera*

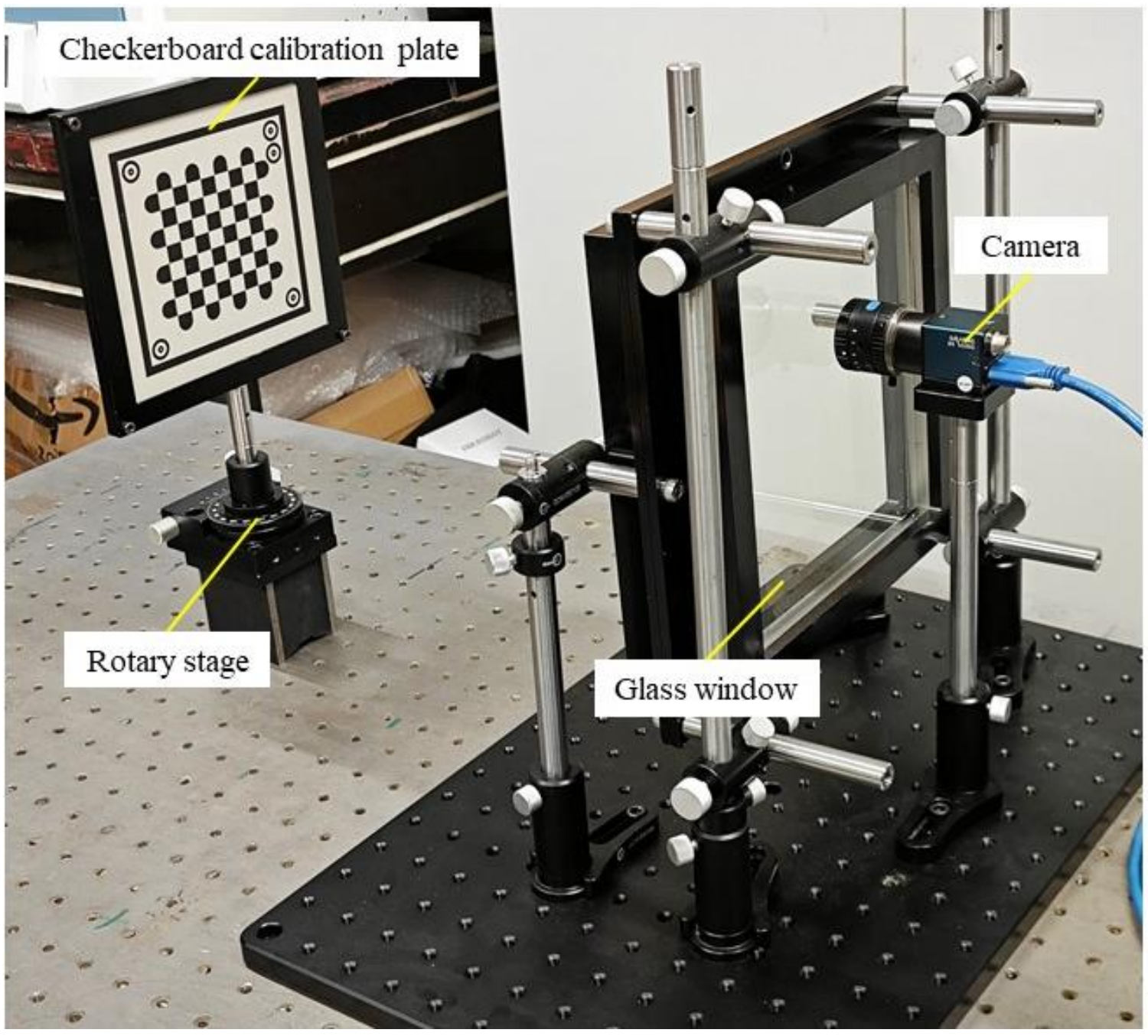

Fig. 2. The experimental setup for single camera calibration.

In order to verify the accuracy of the proposed method, an experimental platform was built as shown in Fig. 2, which mainly consists of a camera, a calibration plate, a glass window, and a rotary stage. The glass window has a thickness of 50mm and is arranged in front of the camera to simulate the refraction situation faced in actual measurements. The rotary stage is used to control the rotation of the calibration plate to meet different pose requirements during calibration. The specific parameters of the camera and glass window are shown in Table 1.

**Table 1. Related parameters of camera and refractive interface**

| | |
|---|---|
| Camera Parameters | |
| Type | Daheng MER2-503-36U3M |
| Focal length | 12 mm |
| Resolution ratio | 2448×2048 |
| Pixel pitch | 3.45×3.45μm |
| Interface Parameters | |
| Material | N-BK7 |
| Thickness | 50 mm |
| Refractive index | 1.5168 |

### *3.2 Orientation estimation of refractive interface*

Five different thicknesses of glass are customized, as shown in Fig. 3, including 10mm, 20mm, 30mm, 40mm, and 50mm, all made of the same material and having the same refractive index as the glass window. When these glasses are stacked seamlessly, the thickness of the refractive interface is considered to have changed. At this point, the imaging position of the calibration point will shift with the change of glass thickness.

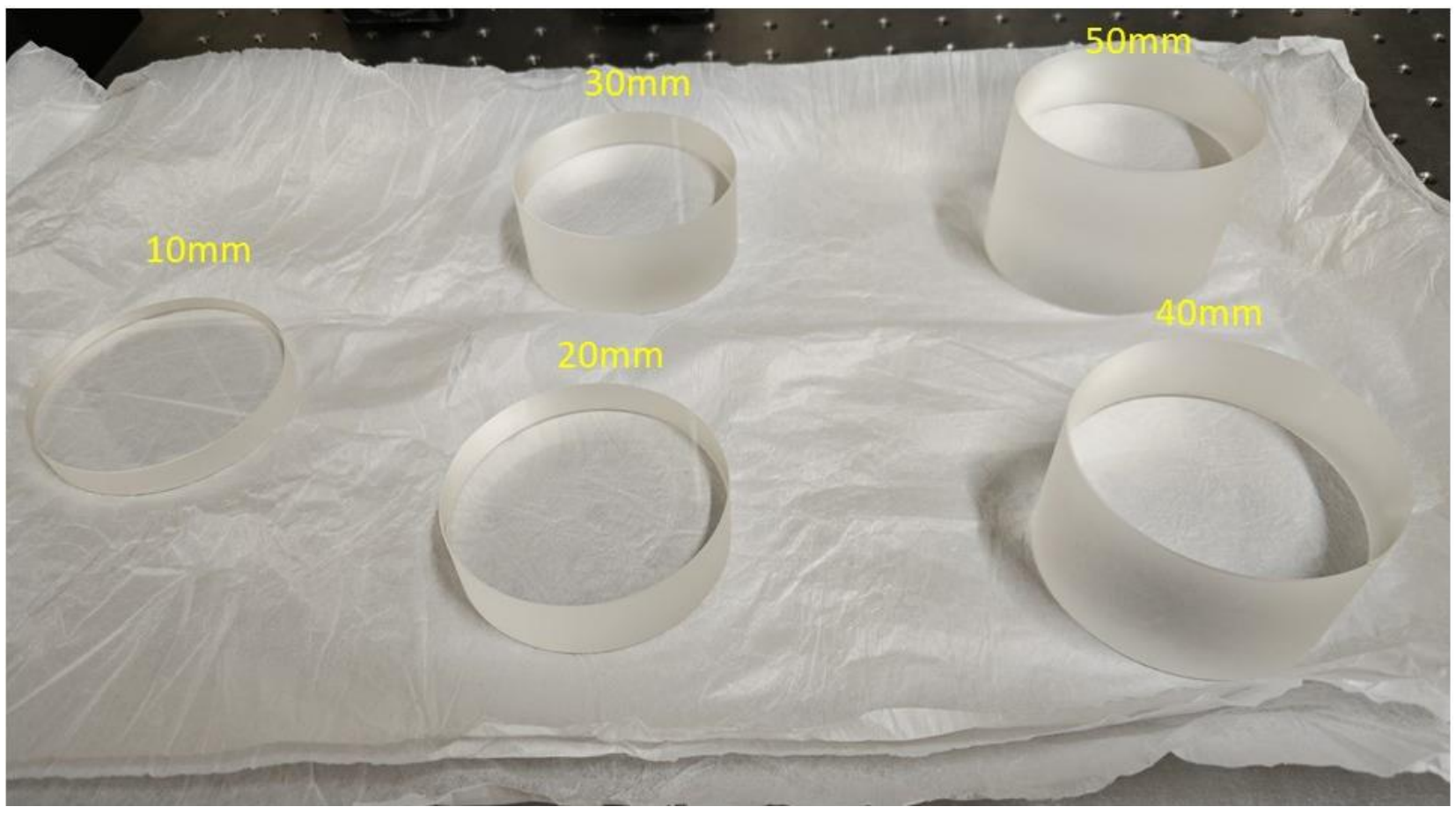

Fig. 3. Five pieces of glass used in the experiment.

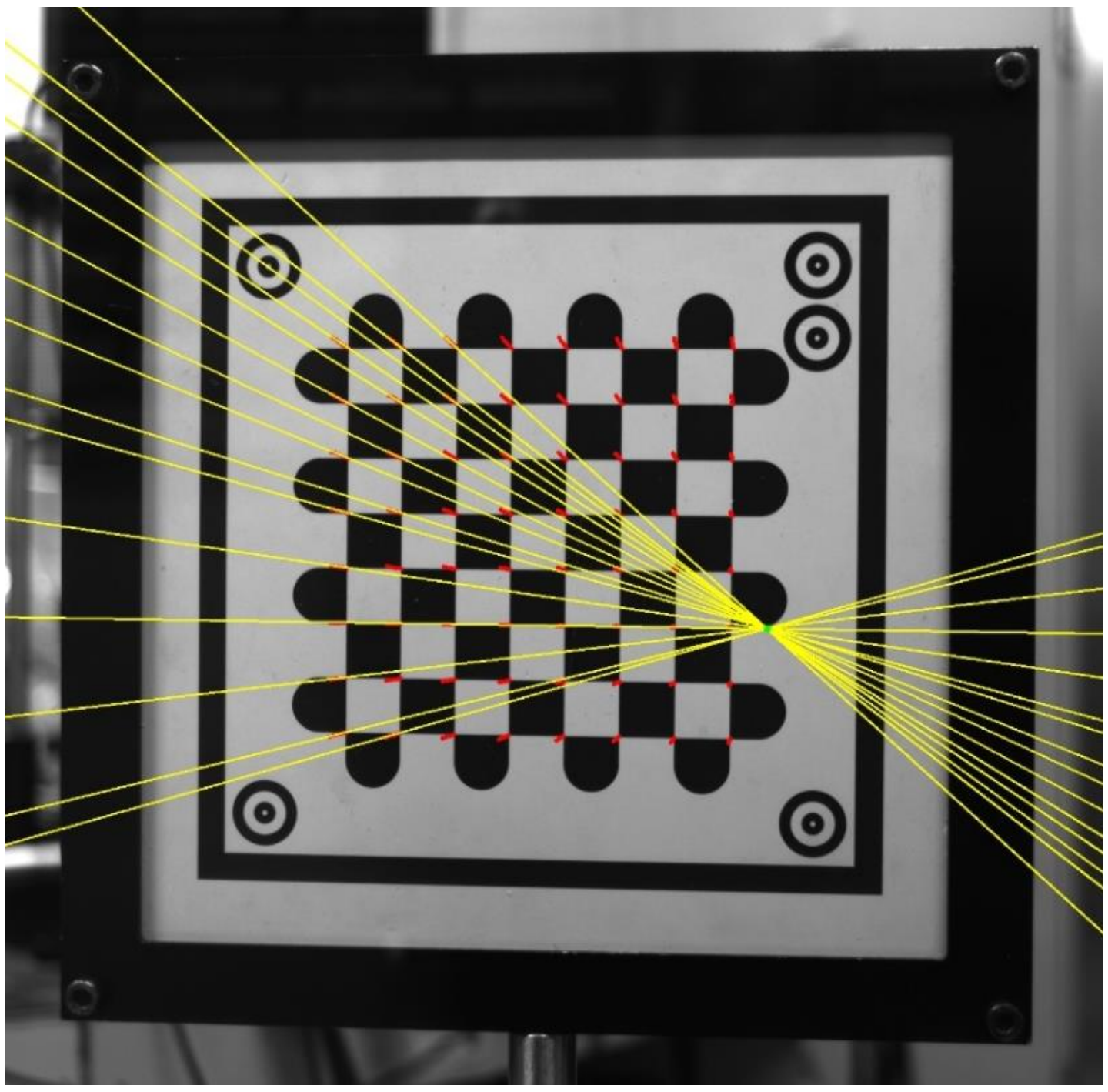

Fig. 4. The fitting result of the expansion center.

Therefore, we captured calibration target images under six different conditions: without glass and with five different glass thicknesses (10-50mm). After extracting corner coordinates from each image, the red points in Fig. 4 were obtained, where each cluster of six adjacent red points represents the displacement trajectory of the same physical corner under varying refractive conditions. These data points were used to fit the movement trajectory of corners as glass thickness changes. Notably, the displacement magnitude varies significantly among different corner positions, which is closely related to their distance from the expansion center - corners closer to this point exhibit smaller displacements. Considering the inevitable measurement errors affecting both x and y directions, we employed the Total Least Squares (TLS) [37] method for trajectory fitting to achieve optimal accuracy. In Fig. 4, the yellow lines clearly demonstrate the fitted trajectories, while the green points accurately indicate the estimated expansion center, showing excellent fitting quality and clarity.

### *3.3 Comparison experiment of calibration methods*

A comparative study was conducted to evaluate the performance of our proposed method against the two most commonly used techniques in refractive imaging - RT [27,38,39] and PF methods [18,40,41].

The RT method follows the procedure proposed by Feng et al. [27]: For each point on the calibration plate, its initial light direction is given by $(-(u-u_0)/f_u, -(v-v_0)/f_v, -1)$, which is determined by the mathematical relationship of light projecting onto the image plane through the optical center. After determining the initial directions, the light path is calculated via ray-tracing and Snell's law until it projects onto the image plane. The reprojection error is obtained by comparing the calculated pixel coordinates with those from the captured image. This error serves as the cost function to optimize the camera's internal and external parameters and the refractive interface related parameters.

The PF method adopts the 3×3×3 parameter form proposed by Paolillo et al. [18], as shown in Eq. (14). This equation requires optimizing forty parameters for each camera, making it the method with the largest number of parameters among the three. The parameter optimization is achieved through the least square method. Notably, the optimization process involves only numerical fitting without physical constraints, so the optimized parameters do not have practical physical meanings.

$$P = \sum_{k=0}^{3} \sum_{j=0}^{3-k} \sum_{i=0}^{3-k-j} a_{ijk} x^i y^j z^k \tag{14}$$

In contrast, the camera calibration method based on the ARI equation proposed in this paper has significant theoretical advantages. This method relies on a rigorous physical optics model, which not only fully accounts for the actual impact of the refractive interface on the imaging process but also endows each parameter in the model with clear physical meanings, demonstrating stronger interpretability and theoretical completeness.

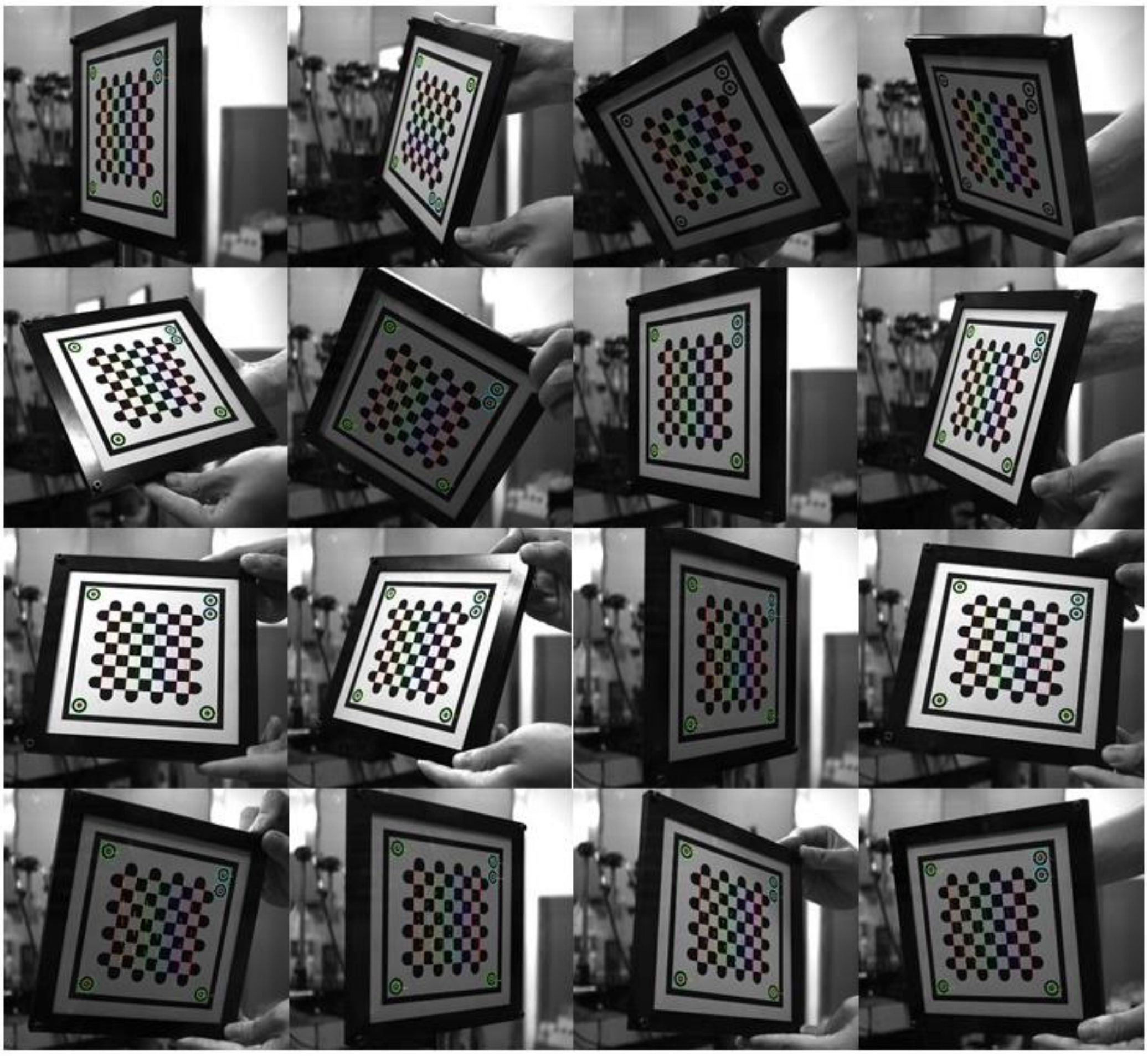

Fig. 5. The posture of the calibration plate and the extraction results of corner points.

By capturing images of calibration plates in different poses, as shown in Fig. 5, three methods were used for camera calibration. Fig. 6 shows the comparison results of reprojection errors among the three calibration methods. From the distribution characteristics of reprojection errors, the error distributions of the three methods are mainly concentrated around zero. Specifically, The RT method has the widest error distribution range, indicating the least ideal calibration effect. This phenomenon stems from two key factors: First, this method requires optimization in a unified coordinate system, and the presence of the refractive interface reduces the accuracy of the transformation matrix between the camera coordinate system and the world coordinate system; second, the optimization process is prone to falling into local minima rather than global optimal solutions and is sensitive to initial errors. Although the PF method achieves the smallest reprojection error, it is essentially an implicit optimization based on mathematical approximation, lacking clear physical meaning. The method proposed in this paper has a slightly larger reprojection error distribution range than the PF method, but this method is based on a clear physical model that can ensure interpretability of the fitting parameters.

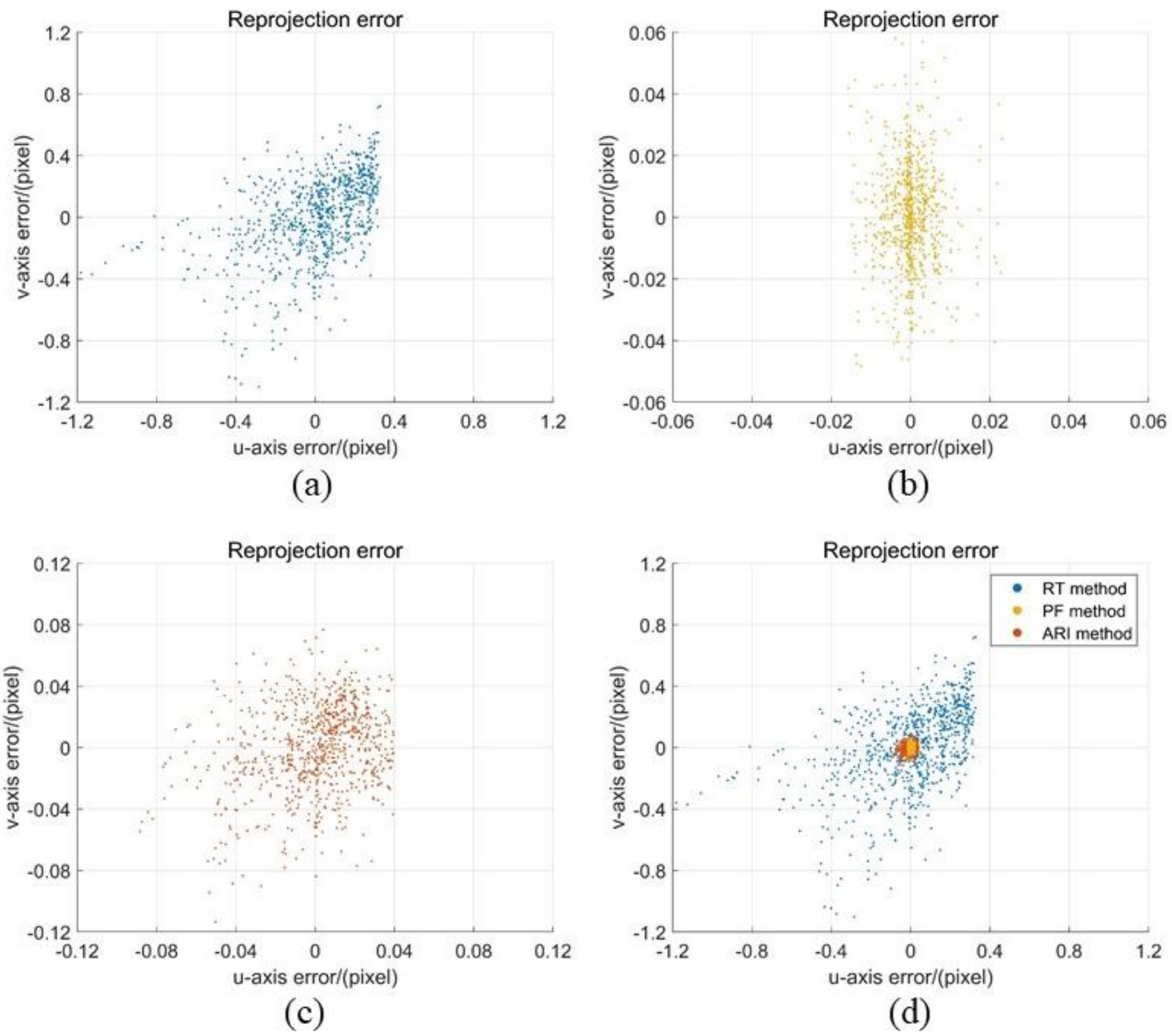

Fig. 6. The reprojection errors of the calibration results of a single camera by three methods. (a) Ray tracing method. (b) Polynomial fitting method. (c) Method base on the analytical refractive imaging equation. (d) Comparison chart of three methods.

**Table 2. Calibration results of single cameras by three methods**

| Method | # of para | MAE(pixels) | Max(pixels) | RMSE(pixels) | T(ms) |
|---|---|---|---|---|---|
| RT | 13 | 0.31 | 1.42 | 0.37 | 298 |
| PF | 40 | 0.01 | 0.08 | 0.02 | 7 |
| ARI | 13 | 0.03 | 0.12 | 0.04 | 10 |

Table 2 presents a quantitative comparative analysis of the performance metrics associated with the three methods under consideration. The primary evaluation parameters include: the number of parameters to be optimized, average absolute error (MAE), maximum error, root mean square error (RMSE), and computation time. The experimental results indicate that, in terms of computational efficiency, the PF method demonstrates advantages owing to its implicit optimization properties. As an explicit optimization approach, the computational efficiency of the ARI method is only marginally lower than that of the PF method; both methods exhibit performance that is over an order of magnitude faster than that of the RT method. In terms of calibration accuracy, all error indicators of the ARI method are only 10% of those of the RT method. Furthermore, it is noteworthy that ARI maintains an accuracy level at around 50% relative to that achieved by the PF method. This comprehensive performance comparison robustly substantiates that the ARI method proposed in this paper successfully strikes a balance between computational efficiency and calibration accuracy while ensuring physical interpretability.

## 4. Binocular camera calibration comparison experiment

### 4.1 Binocular camera calibration

By rotating the calibration plate, a unified coordinate system between cameras can be established [31], which can extend the ARI method to dual camera and multi camera calibration. The experimental setup is shown in Fig. 7. And the calibration process is consistent with the single camera method. Fig. 8 shows the newly defined world coordinate system and the point cloud generated by the rotation of the calibration plate.

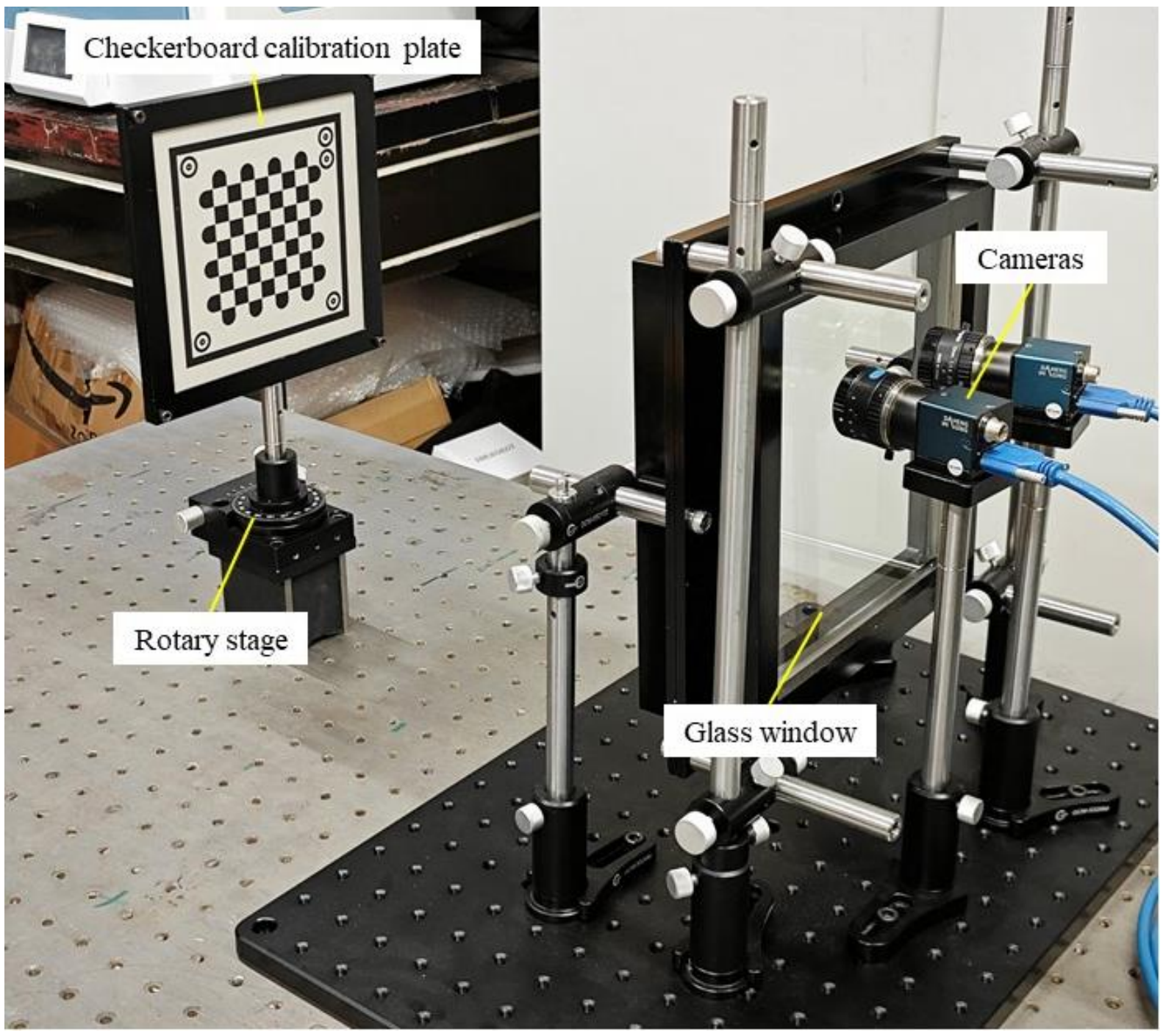

Fig. 7. The experimental setup for binocular camera calibration.

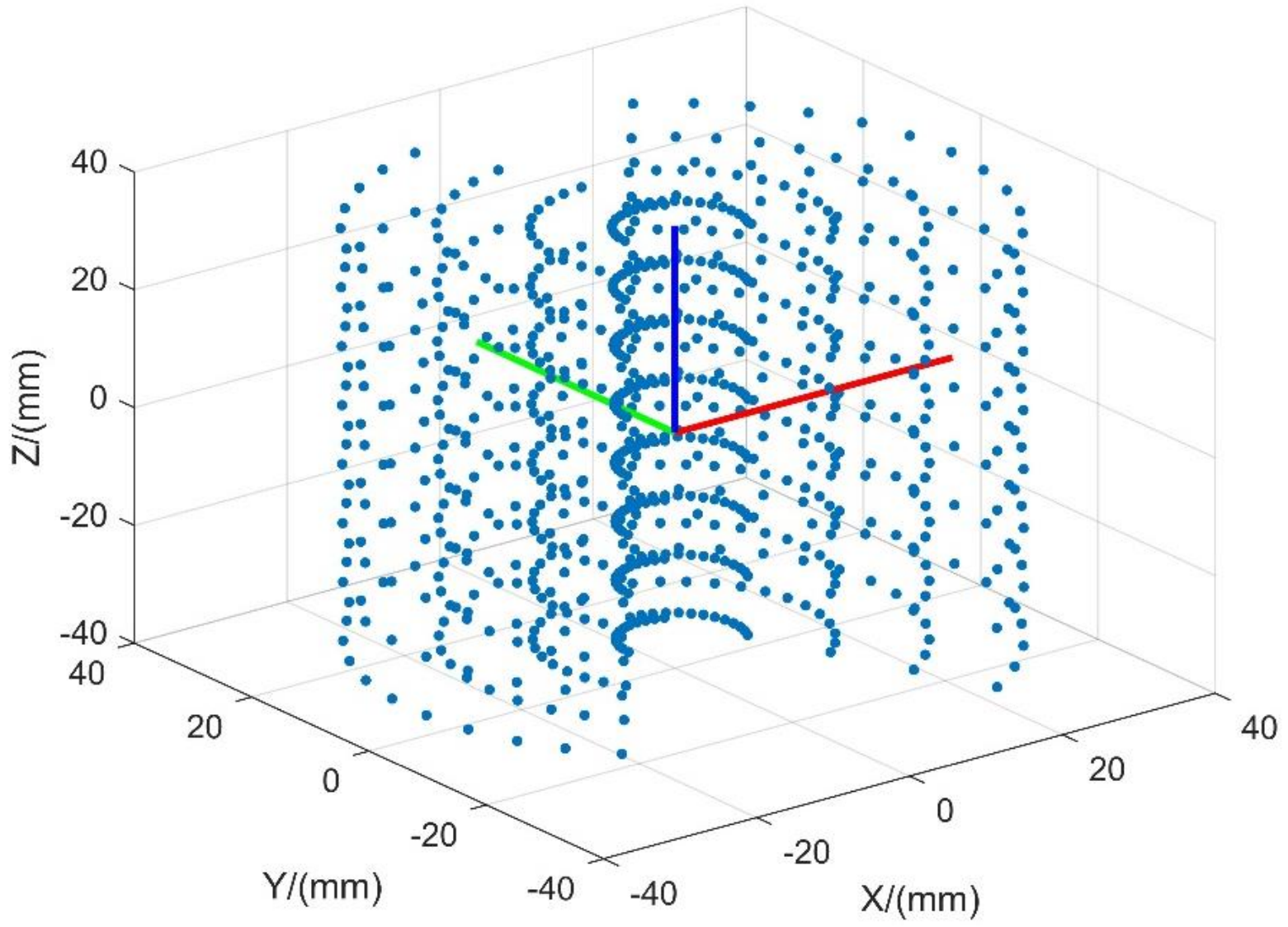

Fig. 8. New world coordinates and corner distribution obtained by rotating the calibration plate.

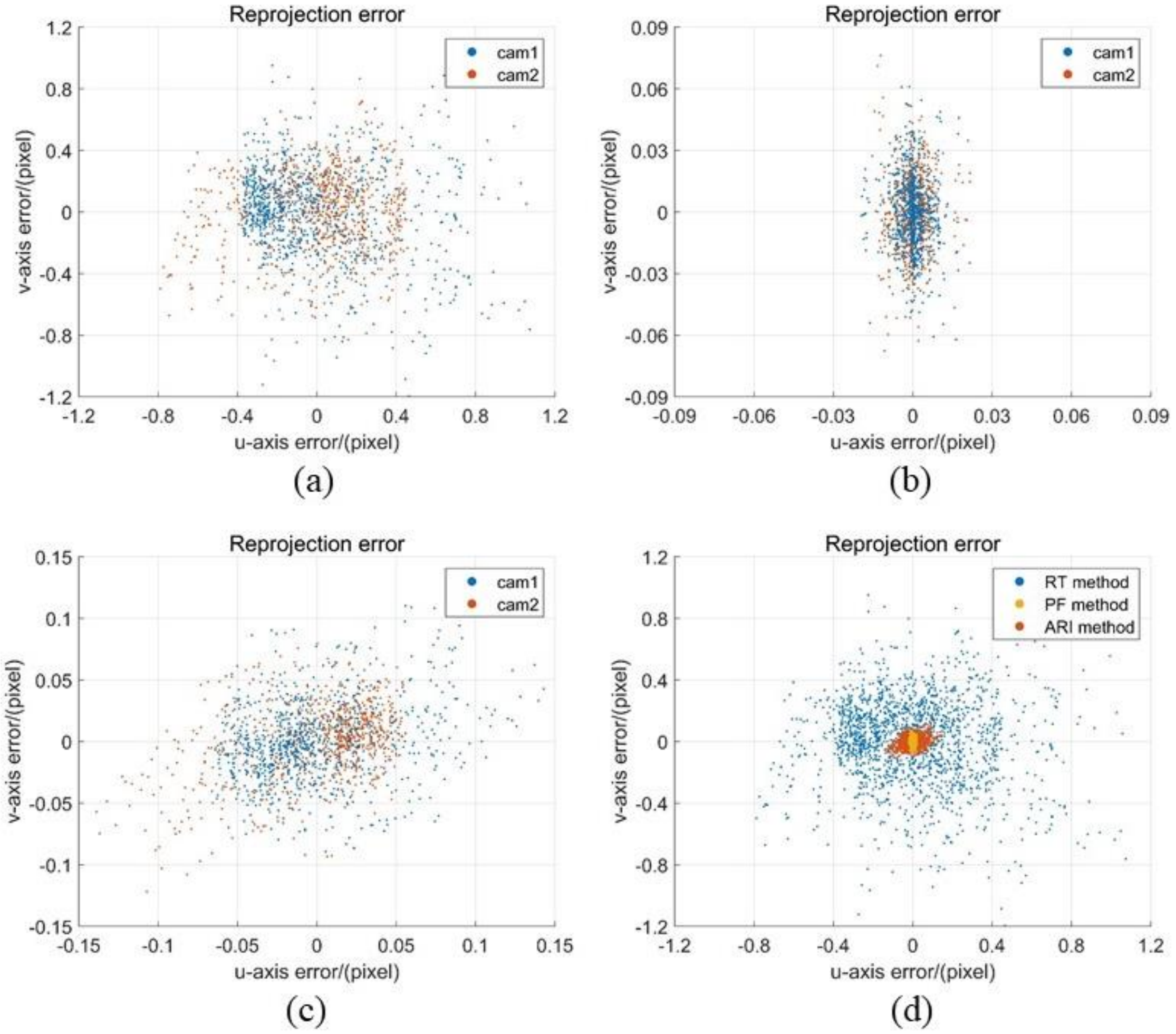

Fig. 9. The reprojection errors of the calibration results of dual cameras by three methods. (a) Ray tracing method. (b) Polynomial fitting method. (c) Method base on the analytical refractive imaging equation. (d) Comparison chart of three methods.

**Table 3. Calibration results of dual cameras by three methods**

| Method | # of para | MAE(pixels) | Max(pixels) | RMSE(pixels) | T(ms) |
|---|---|---|---|---|---|
| RT | 13 | 0.35 | 1.39 | 0.41 | 527 |
| PF | 40 | 0.01 | 0.08 | 0.02 | 12 |
| ARI | 13 | 0.04 | 0.16 | 0.05 | 15 |

The calibration results of the binocular camera based on the point cloud in Fig. 8 are shown in Fig. 9 and Table 3. Overall, the calibration results of the two cameras have good symmetry, and the comparison between various methods is consistent with the calibration results of a single camera.

### *4.2 The limitation of polynomial fitting method*

However, due to the lack of physical model support, the PF method may have hidden dangers in the fitting results when the calibration board cannot cover the observed area. In response to this phenomenon, this study designed a fitting experiment based on partial data. The experiment selected a 3 × 4 array (12 corner points in total) from the center area of the calibration plate for each posture in the binocular camera calibration as the optimization data, and its spatial distribution is shown in Fig. 10. Calculate the reprojection error of all corner points using the optimized parameters from this subset of data, as shown in Fig. 11, and compare the specific values in Table 4. It is worth noting that due to the significant difference in accuracy and efficiency between the RT method and the other two methods in the complete dataset experiment, this comparative experiment only includes the PF method and ARI method.

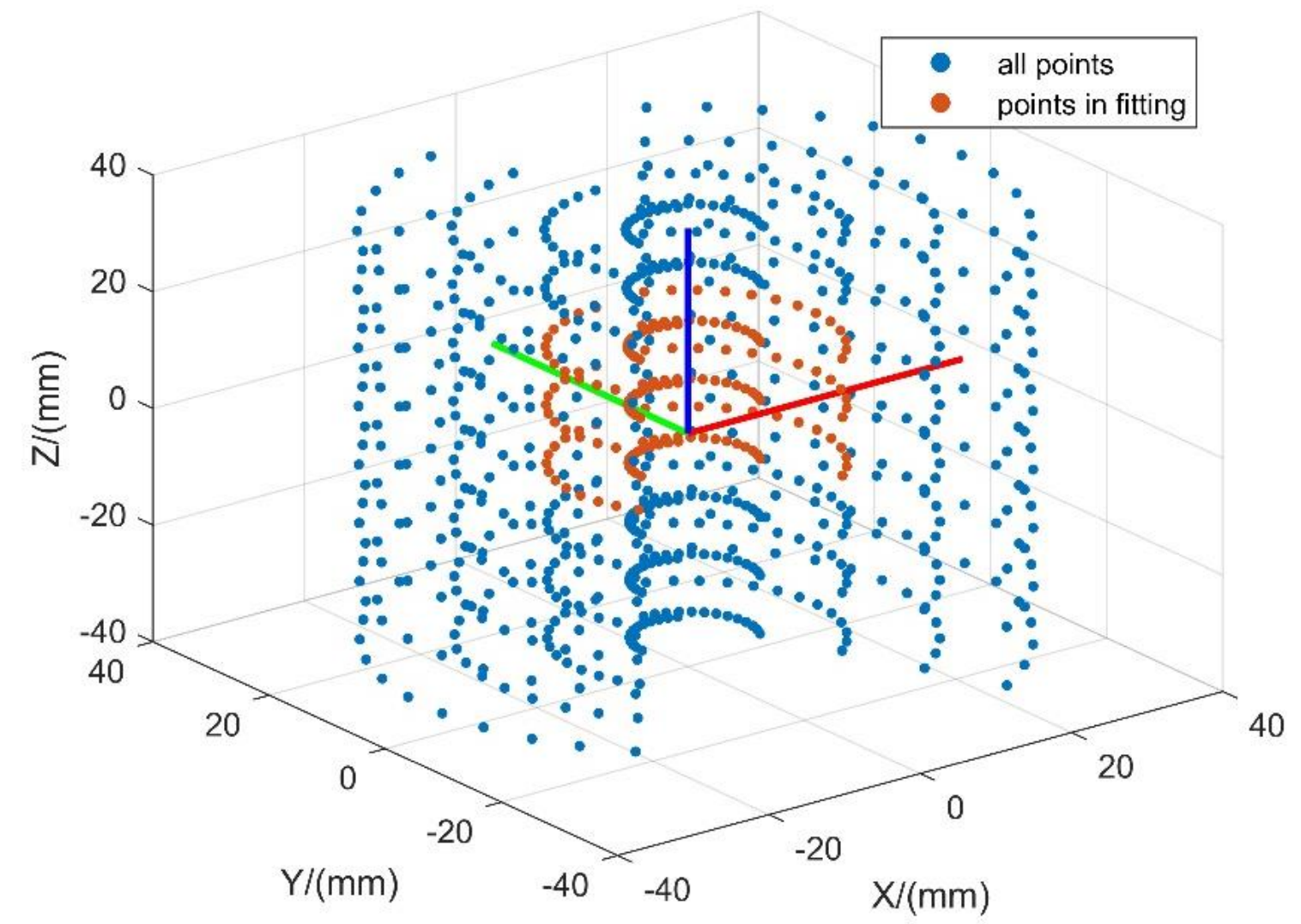

Fig. 10. Distribution of selected corner points in limited data fitting experiments.

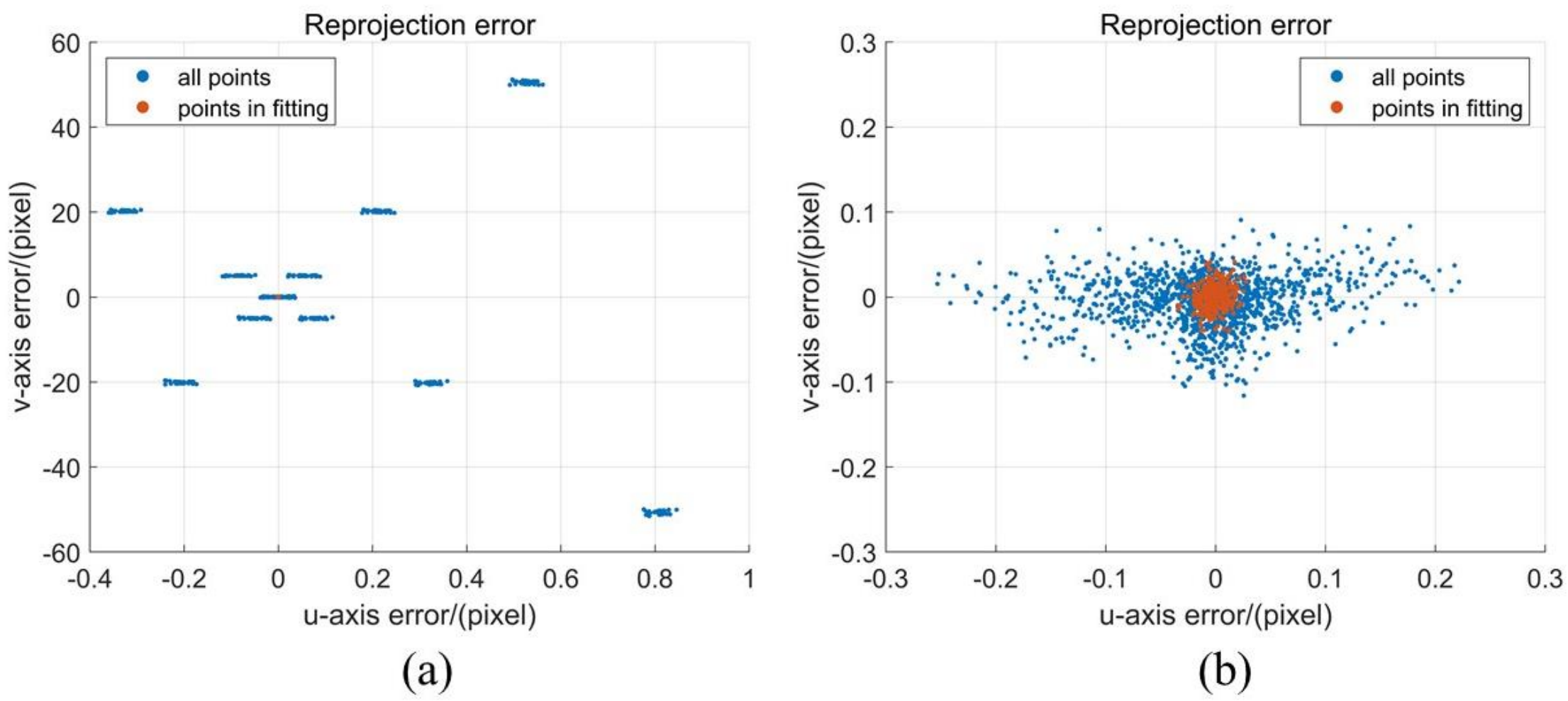

Fig. 11. Reprojection error of limited data fitting experiment. (a) Polynomial fitting method. (b) Method base on the analytical refractive imaging equation.

**Table 4. Performance comparison of two methods in limited data fitting experiment**

| Method | # of para | MAE(pixels) | Max(pixels) | RMSE(pixels) | T(ms) |
|---|---|---|---|---|---|
| PF | 40 | 12.67 | 51.57 | 20.74 | 6 |
| ARI | 13 | 0.05 | 0.25 | 0.07 | 7 |

In Fig. 10, the red markers represent the calibration points involved in parameter optimization, while the blue points denote those not included in the optimization. Correspondingly, Fig. 11 uses red dots to show the reprojection error distribution of the optimized points and blue dots to display the error conditions of the non-optimized points. The experimental results indicate that the calibration points participating in the optimization maintain a very low error level in both methods. However, significant differences emerge between the two methods for the points not involved in the optimization: The error distribution range of outliers in the PF method reaches ±60 pixels, exhibiting an obvious discrete distribution. In contrast, the ARI method keeps the errors highly concentrated, with values remaining close to those of the optimized points.

The statistical data in Table 4 shows that under limited data conditions, the performance of the PF method deteriorates sharply: the MAE soars to 12.67 pixels, and the maximum error reaches 51.57 pixels—approximately 1000 times higher than that with complete data. In contrast, the ARI method only exhibits a 20% increase in MAE and a 35% increase in RMSE, demonstrating excellent robustness. Considering the inherent limitation in practical applications that calibration plates cannot fully cover the target area, the PF method poses significant engineering risks. The ARI method, however, maintains stable calibration accuracy across different scenarios, making it highly valuable for engineering fields such as industrial inspection and underwater observation.

## 5. Binocular vision-based refractive imaging reconstruction experiment

To verify the calibration accuracy, this study conducted reconstruction experiments, and the experimental setup is shown in the Fig. 12. Drive the calibration plate to move back and forth by 10mm using an electrically controlled translation stage, capture an image every 2mm, and use the results of dual target calibration for 3D reconstruction to obtain the corner coordinates of the calibration plate. The reconstruction process adopts the following methods: calculate the projection matrices of two cameras based on optimized parameters, establish corresponding ray equations using the projection matrices and corner pixel coordinates, and determine the three-dimensional coordinates of the corner points by solving the intersection points of the dual camera rays. Considering the situation where actual measurement errors lead to incomplete intersection of rays, the least squares method is used to optimize the solution of spatial point coordinates.

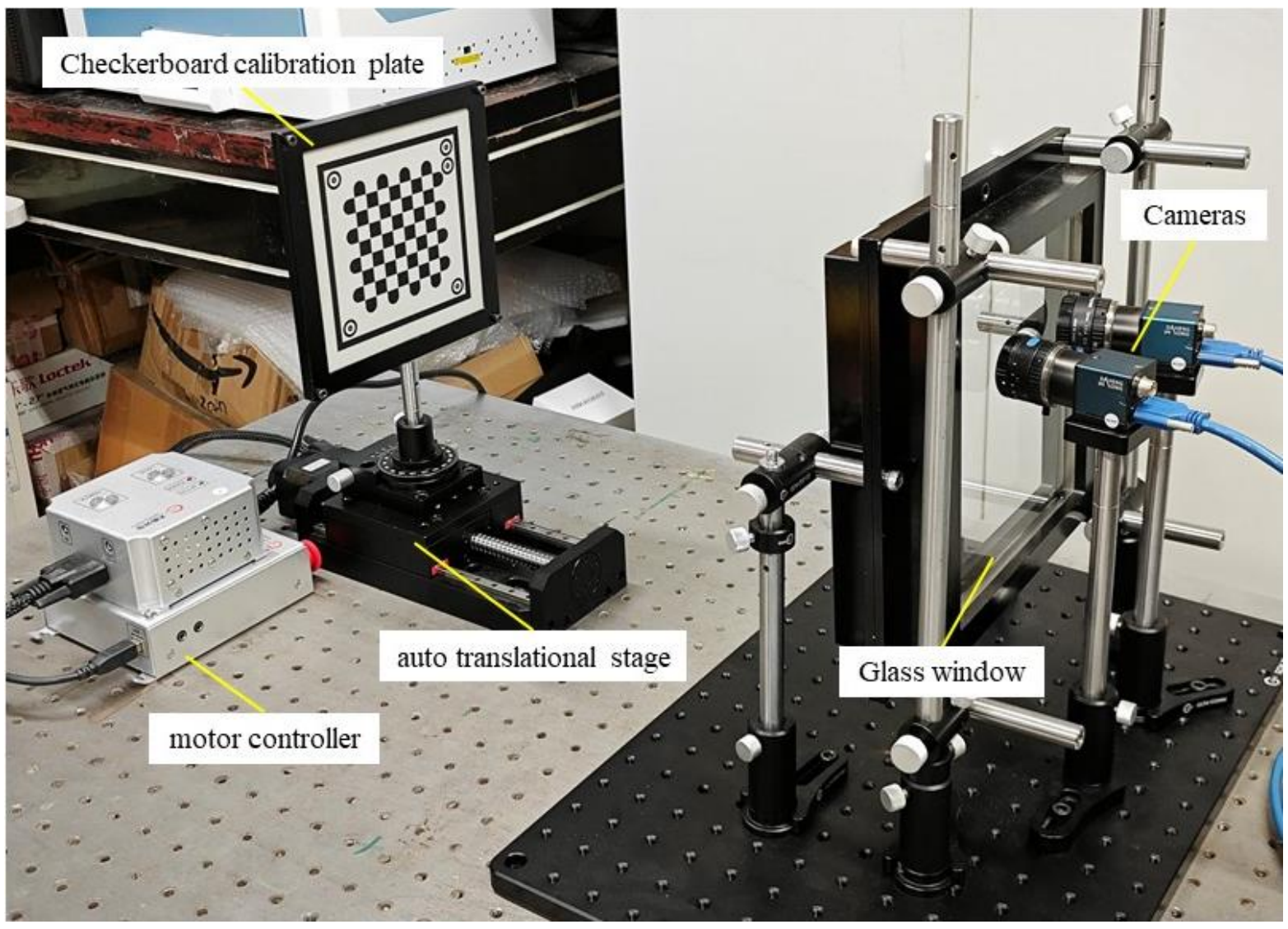

Fig. 12. The experimental setup for reconstruction.

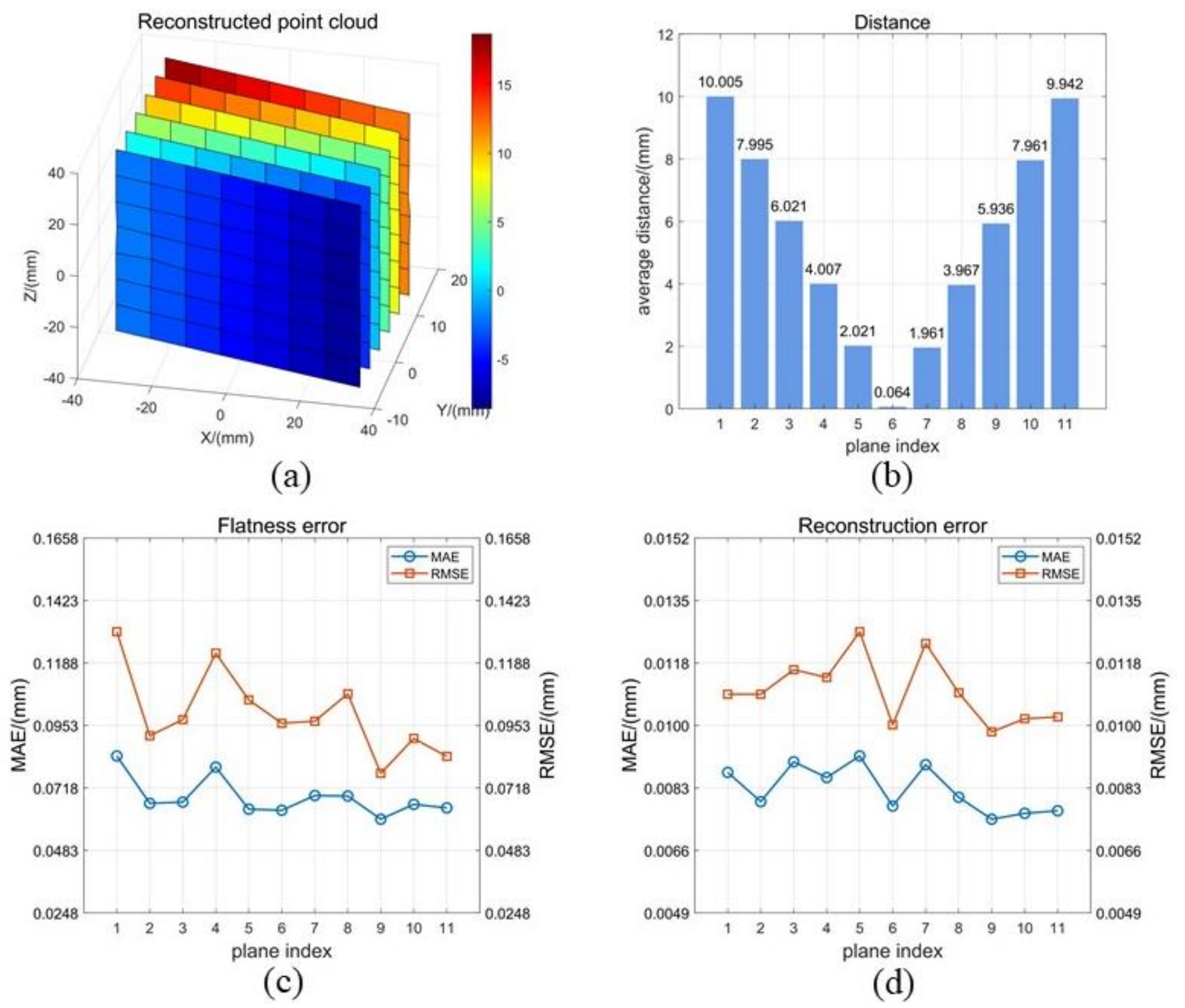

Fig. 13. Reconstruction errors. (a) The reconstructed point cloud distribution. (b) Distance error of reconstructed planes. (c) Flatness error of reconstructed planes. (d) Distance error of reconstructed corner points.

To further discuss the results of the reconstruction, we calculated various parameters of the reconstruction, and the results are shown in Fig. 13. Fig. 13(a) illustrates the distribution of the reconstructed point clouds. For clarity, six representative datasets were uniformly selected from the total eleven groups for visualization. The results demonstrate that the point clouds in each group exhibit well-defined planar distributions with uniform inter-point spacing. Fig. 13(b) further quantifies the distance errors between each group of point clouds and the central plane. The measured errors remain consistently within 0.07 mm, aligning with the experimental design parameters and confirming the reliability of the reconstruction results. Fig. 13(c) presents the distribution of distance errors from the calibration points to the fitted plane. The mean absolute error (MAE) and root mean square error (RMSE) are maintained at 0.07 mm and 0.10 mm, respectively, indicating excellent planarity of the reconstructed point clouds, which closely matches the physical characteristics of the calibration board. Fig. 13(d) displays the statistical error distribution of the reconstructed distances between calibration points (the true value is 10mm), with MAE and RMSE reaching 0.008 mm and 0.011 mm, respectively. These quantitative results strongly validate the high precision and repeatability of the reconstruction results.

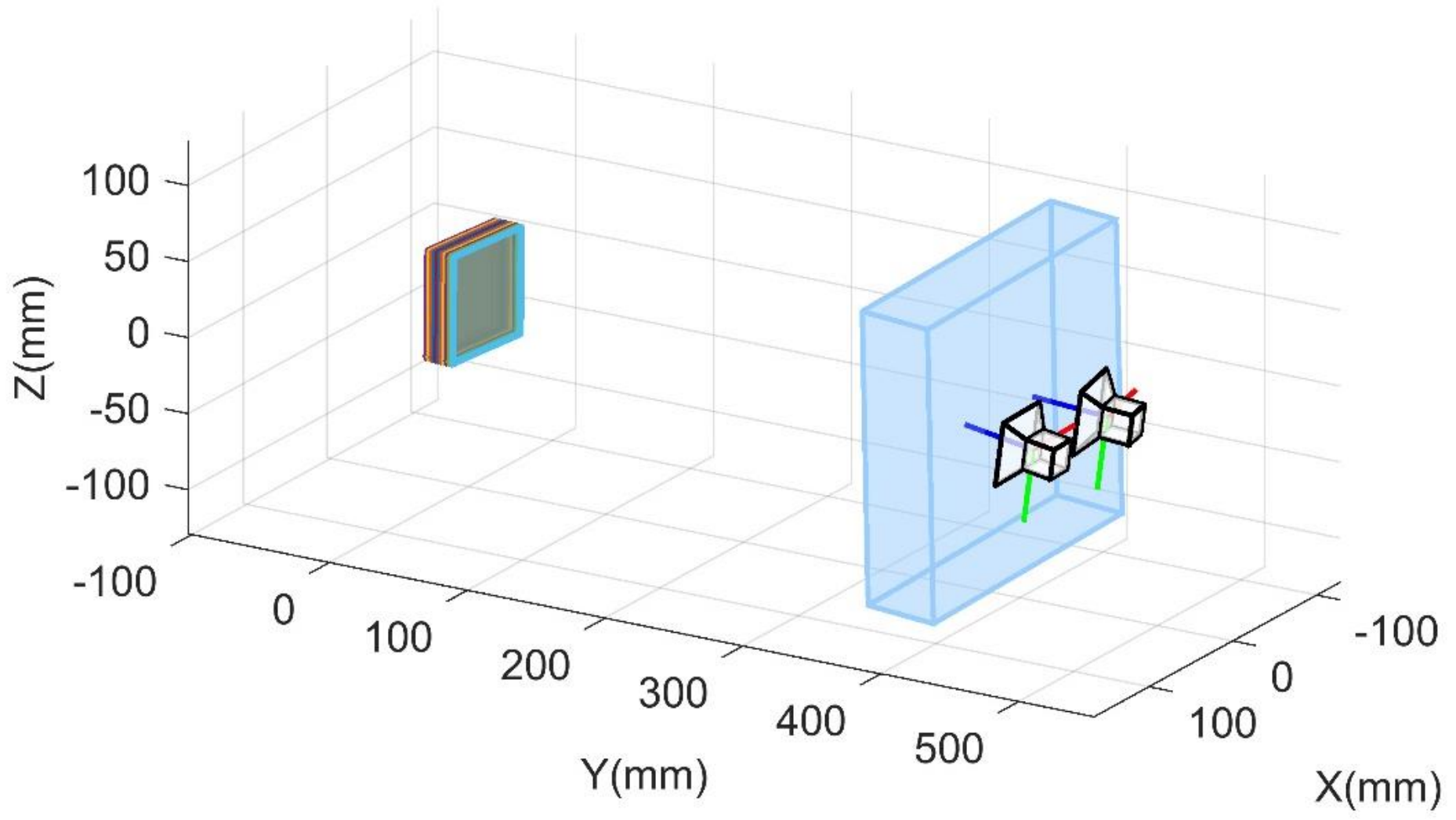

Fig. 14. The result of refractive imaging calibration and point cloud reconstruction

Fig. 14 presents the outcomes of the calibration for both the camera and refraction interface, as well as the reconstruction of the calibration plate utilizing the proposed method. It is evident that the distribution of each component within the entire system aligns with the experimental setup depicted in Fig. 12, thereby further validating the accuracy of the methodology introduced in this article.

## 6. Conclusion

This study presents a novel in-situ joint calibration method for cameras and refractive interfaces, grounded in the ARI equation. This approach eliminates the need for disassembly, enabling simultaneous calibration of both the camera and refractive surfaces solely through glass stacking and rotation of the calibration plate. In comparison to RT method, this technique exhibits approximately 40 times greater computational efficiency while enhancing calibration accuracy by an order of magnitude. Relative to PF approaches, it strictly adheres to physical constraints while maintaining comparable computational efficiency, thereby circumventing the common issue of achieving high local fitting accuracy at the expense of global physical validity. Systematic validation through monocular/binocular calibration experiments and 3D reconstruction tests substantiates the method's superior accuracy and robustness, offering an effective solution for in-situ imaging system calibration within refractive environment.

**Funding.** National Natural Science Foundation of China (62175110); the Funding of NJUST of China (TSXK2022D004).

**Disclosures.** The authors declare no conflicts of interest.

**Data availability.** Data underlying the results presented in this paper are not publicly available at this time but may be obtained from the authors upon reasonable request.